\documentclass{cimento}
\usepackage{pstricks}
\usepackage{color}
\usepackage{amssymb,amsmath,bm,bbm}
\usepackage{epsf,epsfig}
\usepackage{afterpage}
\usepackage{longtable}
\usepackage{cite}
\usepackage{latexsym,mathrsfs,dsfont}
\usepackage{graphics,graphicx,booktabs,array}
\usepackage{url}
\usepackage{paralist}
\usepackage{bbold}

\newcommand{\be}{\begin{equation}}
\newcommand{\ee}{\end{equation}}
\newcommand{\bea}{\begin{eqnarray}}
\newcommand{\eea}{\end{eqnarray}}
\newcommand{\dd}{\displaystyle}
\newcommand{\nn}{\nonumber}
\numberwithin{equation}{section}

\title{Flavour anomalies, correlations, hadronic uncertainties, \\ and all that}
\author{P.~Colangelo\from{ins:x}\thanks{Speaker}\ETC,
F.~De~Fazio\from{ins:x},
F.~Loparco\from{ins:x}
 \atque
N.~Losacco\from{ins:y},\from{ins:x}
}
\instlist{\inst{ins:x} INFN, Sezione di Bari, via Orabona 4,  70125 Bari, Italy
  \inst{ins:y} Dipartimento Interateneo di Fisica "M. Merlin" Universit\`a e Politecnico di Bari, via Orabona 4, 70125 Bari, Italy}

\begin{document}
\maketitle

\begin{abstract}
We present a short overview of the so-called flavour anomalies,  discussing their significance and the connections with  QCD issues discussed at the HADRON 2023 conference.
\end{abstract}

\section{Introduction}
 The Standard Model (SM) of fundamental interactions has achieved an astonishing success,  both in precision and in extension,   in describing the dynamics of the constituents of nature. Nevertheless, there are arguments 
suggesting that it  is not the ultimate theory.  Difficulties concern the neutrino masses and mixings,  the baryon-antibaryon asymmetry in the universe, the  dark matter. There are also conceptual issues to address, as the large number of unrelated parameters of the theory, the dynamics of the scalar sector and  its impact on the fermion and gauge sectors, the arbitrariness  of the quantum numbers assignment, not to say the connection with gravity. Many difficulties are  rooted in the flavor sector of the model, where  a few anomalies have emerged recently. The implications and perspectives of such anomalies deserve a discussion. They are connected  with  issues discussed at  this conference, since the control of the hadronic uncertainties is of prime importance to assess the significance of the observed deviations.

\section{A list of tensions between  SM expectations and measurements}

\smallskip
\noindent {\bf Lepton Flavour Universality in exclusive semileptonic $b \to c$ transitions} 
\smallskip

In SM the coupling of the gauge bosons to leptons is the same for all families. The {\it Lepton Flavour Universality (LFU)} property has, among others, the consequence that the ratios of semileptonic  decay rates of $B$ to charmed mesons can be precisely predicted if  the  hadronic  form factors of the $b \to c$ current are efficiently constrained. Measurements of the ratios 
$\dd R_{D^{(*)}} = \frac{\Gamma(B \to D^{(*)} \tau \bar \nu_\tau)}{\Gamma(B \to D^{(*)} \ell \bar \nu_\ell)}$, 
with  $\ell=\mu,e$,  produce a combined result  3.4$\sigma$ away from SM \cite{ParticleDataGroup:2022pth,Gambino:2020jvv,HFLAV:2022pwe}, an outcome of debated interpretation. If the effect is attributed to a beyond the SM (BSM) phenomenon, the
 interpretation is easier in an effective field theory approach, allowing  for $D=6$ operators  not present in the  low-energy SM Hamiltonian:
\bea
H_{\rm eff}^{b \to c \ell \nu}&=& {G_F \over \sqrt{2}} V_{cb} \Big[(1+\epsilon_V^\ell) \left({\bar c} \gamma_\mu (1-\gamma_5) b \right)\left( {\bar \ell} \gamma^\mu (1-\gamma_5) {\nu}_\ell \right) \nn \\
&+& \epsilon_R^\ell \left({\bar c} \gamma_\mu (1+\gamma_5) b \right)\left( {\bar \ell} \gamma^\mu (1-\gamma_5) {\nu}_\ell \right) 
+ \epsilon_S^\ell \, ({\bar c} b) \left( {\bar \ell} (1-\gamma_5) { \nu}_\ell \right) \nn \\
&+& \epsilon_P^\ell \, \left({\bar c} \gamma_5 b\right)  \left({\bar \ell} (1-\gamma_5) { \nu}_\ell \right)
+ \epsilon_T^\ell \, \left({\bar c} \sigma_{\mu \nu} (1-\gamma_5) b\right) \,\left( {\bar \ell} \sigma^{\mu \nu} (1-\gamma_5) { \nu}_\ell \right) \Big] + H.c.\, , \label{hamil} 
\eea
and experimentally constraining  the Wilson coefficients.  In \eqref{hamil}  $G_F$  is the Fermi constant and $V_{cb}$ is the  Cabibbo-Kobayashi-Maskawa (CKM) matrix element. 
$H_{\rm eff}$ involves four-fermion operators with  left-handed neutrinos and complex  lepton-flavour dependent  coefficients $\epsilon^\ell_{i}$. 
 The SM is recovered for  $\epsilon^\ell_{V,R,S,P,T}=0$. 
After constraining the coefficients using the $B$ modes,  related effects are foreseen in different channels for mesons ($B_s, B_c$) and baryons ($\Lambda_b$), exclusive and inclusive.  Observing  correlated deviations from SM would be a smoking gun for the existence of the effect: we  discuss below what is expected for $\Lambda_b$ decays.  The correlations could overcome the uncertainties  in the hadronic matrix elements \cite{Colangelo:2018cnj,Colangelo:2020vhu,Colangelo:2023xnu}. On the other hand, the treatment of the hadronic quantities is advocated  to challenge the new physics (NP) interpretation of this tension  \cite{Martinelli:2023fwm}.

\bigskip
\noindent {\bf Determination of $|V_{ub}|$ and  $|V_{cb}|$ from exclusive and inclusive $B$ decay modes}
\smallskip

In SM the elements of the CKM matrix are parameters to be determined experimentally. 
A long-standing issue  concerning the measurement of  $|V_{ub}|$ and $|V_{cb}|$ is that the most precise determinations, done using B decays, are in tension if inclusive or exclusive modes are exploited \cite{HFLAV:2022pwe}. 
As we shall see below, the inclusive measurements rely on a systematic QCD expansion in the heavy quark mass and in the strong coupling,  also using moments of the lepton energy spectrum  \cite{Bernlochner:2022ucr}. Third order corrections to the moments in $B \to X_c \ell \nu $ have been  considered \cite{Fael:2022frj},  as well as the electromagnetic corrections  \cite{Bordone:2021oof}.  The exclusive 
determinations rely on processes such as $B  \to D^{(*)} \ell \nu$, for which the  uncertainty related to the hadron form factors can be treated invoking arguments based on QCD symmetries, such as the heavy quark symmetry  holding in the large $m_b$ limit \cite{Neubert:1993mb}, producing high-precision results  \cite{Bigi:2017jbd,Bigi:2023cbv,Bigi:2017njr,Bigi:2016mdz,Belle-II:2023okj}. The treatment of the hadronic form factors, including dispersive bounds, can remove the tension  \cite{Martinelli:2023fwm}; differently, a possible connection with the $b \to c$ semileptonic  anomaly has been investigated  \cite{Colangelo:2016ymy}.

\bigskip
\noindent {\bf Unitarity relations in the Cabibbo-Kobayashi-Maskawa matrix}
\smallskip

A deficit has been observed in the unitarity relations involving the elements of the first row and  the first column of the CKM matrix \cite{ParticleDataGroup:2022pth}:
\be
|V_{ud}|^2+|V_{us}|^2+|V_{ub}|^2=0.9985\pm0.0005 , \hspace{0.4cm}
|V_{ud}|^2+|V_{cd}|^2+|V_{td}|^2=0.9970\pm0.0018  \,\,\, . \nn
\ee
Due to the small values of $V_{ub}$ and $V_{td}$, the deficit concerns the first two terms  of the  relations ({\it Cabibbo angle anomaly}). The attention is focused on the determinations of $V_{ud}$ from nuclear $\beta$ decays with the role of the radiative corrections \cite{Czarnecki:2018okw,Hardy:2020qwl,Marciano:2005ec,Seng:2018yzq}, and on the extraction of $V_{us}$ from leptonic and semileptonic $K^+$ decays  \cite{KLOE:2005xes,KLOE:2007wlh,Seng:2021nar,Cirigliano:2022yyo}. Determinations in $\tau$ decays are also scrutinized. The possible origin from BSM phenomena  has been considered  \cite{Belfatto:2019swo}, namely investigating the effects of modified neutrino couplings \cite{Coutinho:2019aiy}.

\bigskip
\noindent {\bf Observables in $b \to s \ell^+ \ell^-$  processes}
\smallskip

Processes as those induced by $b \to s \ell^+ \ell^-$ and the other Flavour Changing Neutral Current (FCNC) transitions, which in SM occur at loop level,  are  sensitive to heavy quanta  contributions.  Anomalies have been detected in decay rates, such  as  $\Gamma(B \to K \mu^+ \mu^-)$ and  $\Gamma(B_s \to \phi \mu^+ \mu^-)$
\cite{LHCb:2014cxe,Parrott:2022zte,LHCb:2021zwz,Gubernari:2022hxn}  and in observables constructed from  the  $B \to K^* \mu^+ \mu^-$ angular distributions   \cite{Descotes-Genon:2012isb,LHCb:2020lmf}.
  Also semi-inclusive transitions show  difficulties \cite{Isidori:2023unk}.  On the other hand, the $B_s \to  \mu^+ \mu^-$ decay rate is compatible with SM \cite{Buras:2022qip}.
An enhancement of the rate of another FCNC process, $B \to K \nu \bar \nu$ with respect to the SM expectation has been  recently observed by the Belle-II Collaboration
\cite{Belle-II:2023esi}, 
a result  requiring  a dedicated discussion.

\bigskip
\noindent {\bf Anomalous magnetic moment of the muon} 
\smallskip

The measurement  in \cite{Muong-2:2023cdq}  combined with the  results  in \cite{Muong-2:2021ojo,Muong-2:2006rrc} has provided a determination  of the muon anomalous magnetic moment $a_\mu={(g-2)_\mu}/{2}$ with the  precision of 0.20 ppm. It deviates  from the SM result  quoted in the White Paper \cite{Aoyama:2020ynm}:
\be 
a_\mu^{exp}=116 \, 592 \, 059 \, (22) \times 10^{-11} \quad, \hspace{0.8cm} a_\mu^{WP}=116 \, 591 \, 810 \, (43) \times 10^{-11}  , \nn
\ee
a tension the origin of which is under scrutiny.  Improvement in the measurement is foreseen at Fermilab. The main uncertainty in the SM determination is in the hadronic contributions to $a_\mu$. New evaluations of the hadronic light-by-light contributions, e.g. in \cite{Colangelo:2023een}, confirm the previous results obtained by dispersive analyses  (see \cite{Colangelo:2023rqr} and  the references therein). The hadronic vacuum polarization (HVP) contribution is determined by  dispersive analyses with the measured $e^+ e^- \to {\rm hadron}$ cross section as an input \cite{Aoyama:2020ynm}. A value of $a_\mu$ more consistent with experiment is obtained in a lattice QCD computation of the HVP contribution \cite{Borsanyi:2020mff}.   A tension is found in the comparison of  the $R$ ratio measurements  in the low-s range with lattice QCD determinations \cite{ExtendedTwistedMassCollaborationETMC:2022sta}. Moreover, a  measurement of the $e^+ e^- \to \pi^+ \pi^-$ cross section from threshold to 1.2 GeV disagrees with previous results \cite{CMD-3:2023alj}. The  situation is intriguing,  improved analyses  are ongoing, but the
 issue  stresses the role of controlling the hadronic effects in  precision observables.

\section{Interplay between flavour sector  and  hadron physics}
As we have seen, there is an interplay between the tensions in SM observables and the hadron physics discussed at this conference. A few new examples are given below.

\bigskip
\noindent {\bf Inclusive semileptonic and radiative decays of $b$ baryons} 
\smallskip

\begin{figure}[t!]
\begin{center}
\includegraphics[width = 0.4\textwidth]{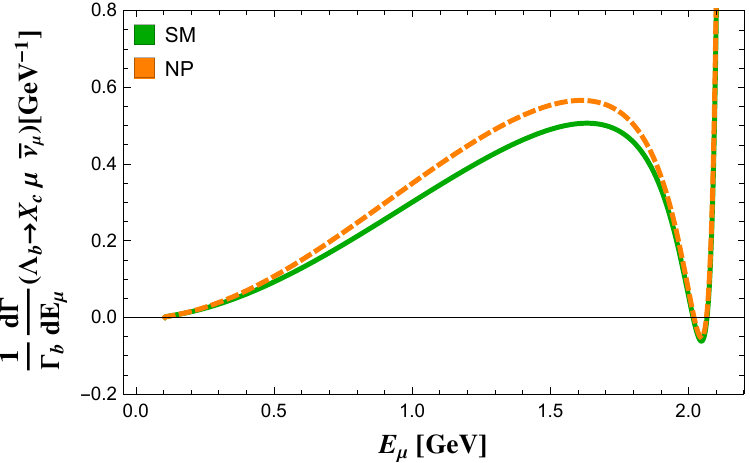} 
\hskip 0.3cm 
\includegraphics[width = 0.4\textwidth]{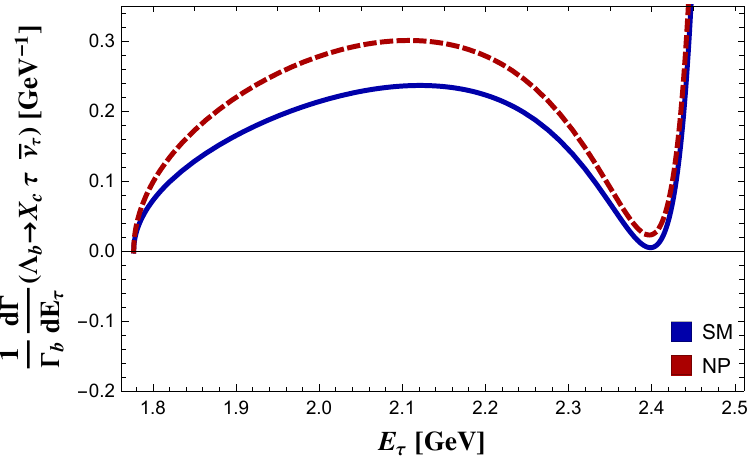} \\ \vskip 0.3cm
\includegraphics[width = 0.4\textwidth]{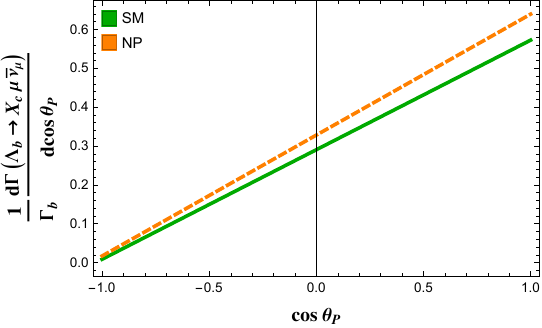} 
\hskip 0.3cm 
\includegraphics[width = 0.4\textwidth]{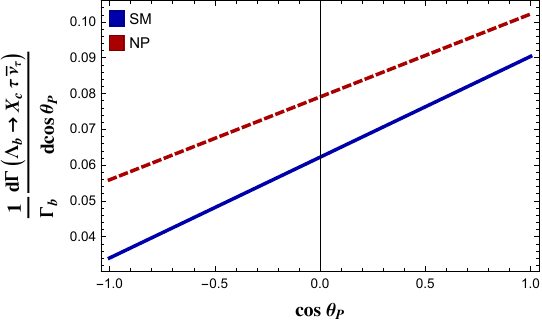} 
\caption{ \baselineskip 12pt  \small Charged lepton energy spectrum (top panels)  and $\dd \frac{1}{\Gamma_b} \frac{d \Gamma}{d\cos \theta_P}$ distribution (bottom panels) for $\Lambda_b \to X_c \ell \bar \nu_\ell$, with $\ell=\mu$ (left) and  $\ell=\tau$ (right).
 The solid line is the SM result, the dashed line the result for NP at a benchmark point for the Wilson coefficients compatible with $B \to D^* \ell \nu_\ell$ data  \cite{Colangelo:2018cnj,Colangelo:2020vhu}.  }\label{fig:ElspectrumCharm}
\end{center}
\end{figure}

The  inclusive $b \to c$ semileptonic decay  of a baryon $H_b$ comprising a single $b$ quark
\be
H_b(p,s) \to X_{c}(p_X) \ell^-(p_\ell) {\bar \nu_\ell}(p_\nu) \,\,\, ,  \label{inclusive-decay}
\ee
with $s$  the spin of the decaying baryon,  is induced by the generalized  low-energy  Hamiltonian \eqref{hamil}  written as
\be
H_{\rm eff}^{b \to c \ell \nu}= {G_F \over \sqrt{2}} V_{cb} \sum_{i=1}^5 C_i^\ell \, J^{(i)}_M\, L^{(i)M} + H.c.\,\,\, .\label{hnew}\ee
   $J_M^{(i)}$ is  the hadronic and $L^{(i)M}$ the leptonic current in each operator, with $M$  the set of Lorentz indices contracted between $J$ and $L$. 
The  decay width of  \eqref{inclusive-decay}  is obtained from
\be
d\Gamma= d\Sigma \, \frac{G_F^2 |V_{cb}|^2}{4m_H} \sum_{i,j} C_i^* C_j (W^{ij})_{MN} (L^{ij})^{MN} 
\ee
with  phase-space $d\Sigma$. By the optical theorem, the hadronic tensor $(W^{ij})_{MN}$ is related to
 the discontinuity of the forward  amplitude
\bea
(T^{ij})_{MN}&=&i\,\int d^4x \, e^{-i\,q \cdot x} \langle H_b(p,s)|T[ J^{(i)\dagger}_M (x) \,J^{(j)}_N (0)] |H_b(p,s) \rangle\,\,\label{Tij-gen}
\eea
 across the cut corresponding to the process \eqref{inclusive-decay}.
This can be computed  by an operator product expansion (OPE)  in the inverse $b$ quark mass  \cite{Chay:1990da,Bigi:1993fe}.
The resulting expression involves $H_b$ matrix elements of QCD operators of increasing dimension,
\be
{\cal M}_{\mu_1 \dots \mu_n}=\langle H_b(v,s)|({\bar b}_v)_a(i D_{\mu_1})\dots(i D_{\mu_n})(b_v)_b |H_b(v,s)\rangle \label{matel}
\ee  
($a,b$  Dirac indices),  given  in terms of  nonperturbative parameters, the number of which increases with the dimension of the  operators.
 The matrix elements needed for the expansion at ${\cal O}(1/m_b^3)$ keeping the  $s_\mu$ dependence are given in \cite{Colangelo:2020vhu}. With such expressions one can compute, e.g.,  the distributions in the charged lepton energy and  in the angle  $\theta_P$  between ${\vec p}_\ell$ and ${\vec s}$  in the $H_b$ rest frame. In  Fig.~\ref{fig:ElspectrumCharm} the deviation correlated to the anomaly in semileptonic $B\to D^{(*)} \tau \nu_\tau$  decays can be appreciated \cite{Colangelo:2018cnj,Colangelo:2020vhu}. $\Lambda_b$ with sizable polarization are expected to be produced at the new lepton colliders, with the $b$ quarks coming from $Z^0$ and top quark decays.
 Inclusive $b \to u$  semileptonic  modes  and  rare radiative modes can be described by the same approach. The treatment of the singular terms in the distributions involves a nonperturbative shape function,  a new expression of which  has been derived  in \cite{Colangelo:2023xnu}.

\bigskip
\noindent{\bf Exclusive $B_c$  decays to charmonium}
\smallskip

\begin{figure}[t]
\begin{center}
\includegraphics[width = 0.35\textwidth]{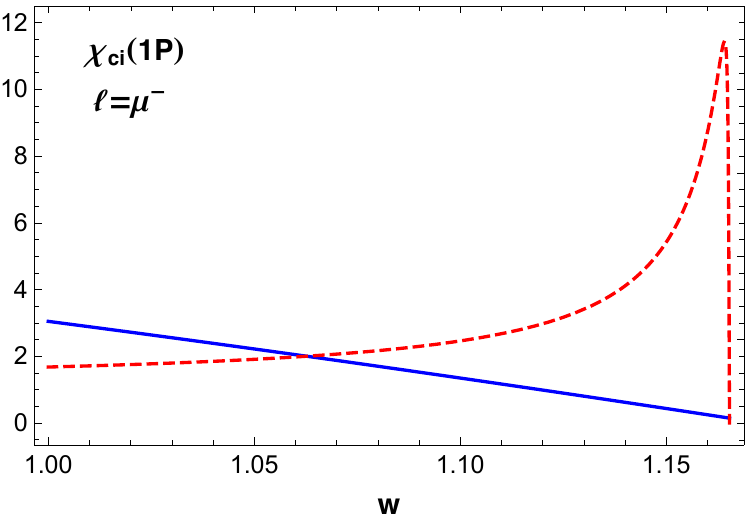}\hskip 0.35cm \includegraphics[width = 0.35\textwidth]{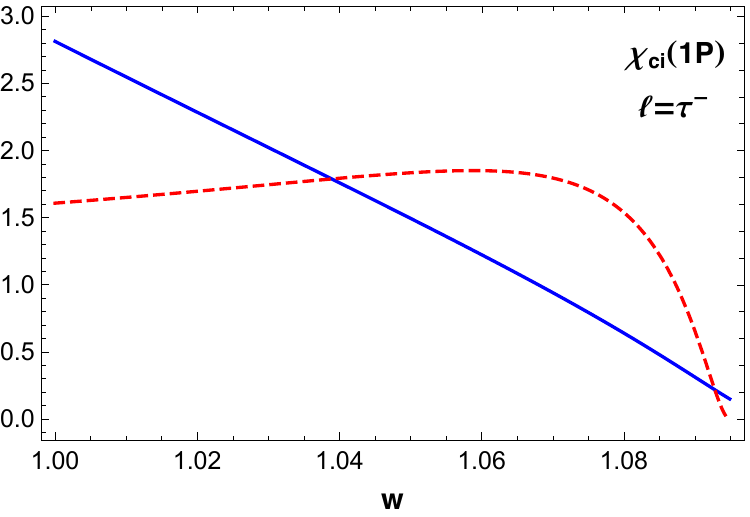}\\
\includegraphics[width = 0.35\textwidth]{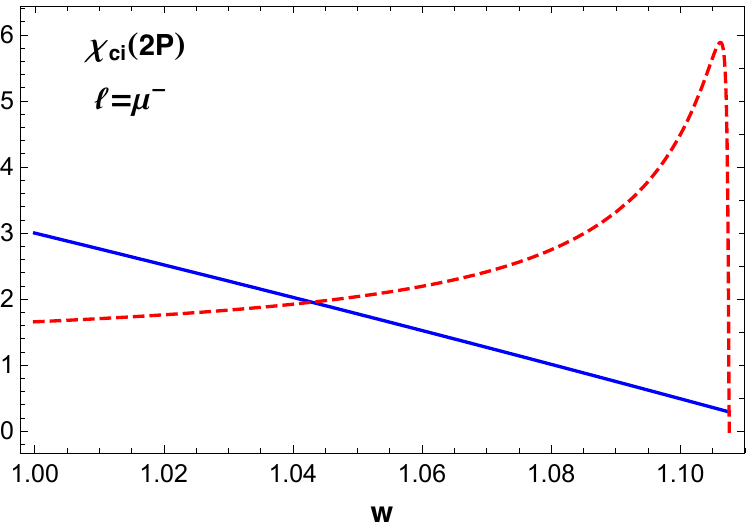}\hskip 0.35cm \includegraphics[width = 0.35\textwidth]{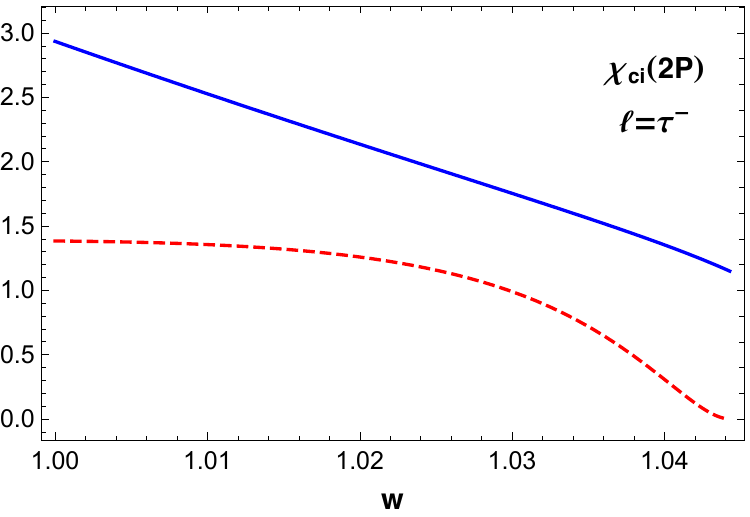}
    \caption{\baselineskip 12pt  \small  Ratios of distributions  $\dd \frac{d \Gamma(B_c \to \chi_{c1} \ell \bar \nu)/dw}{d \Gamma(B_c \to \chi_{c0} \ell \bar \nu)/dw}$ (solid lines) and 
    $\dd \frac{d \Gamma(B_c \to \chi_{c2} \ell \bar \nu)/dw}{d \Gamma(B_c \to \chi_{c1} \ell \bar \nu)/dw}$  (dashed lines) in SM, for $\ell=\mu$ (left) and $\ell=\tau$ (right), and  for the $1P$ (top panels) and $2P$ final charmonia (bottom panels). The meson masses  are quoted in \cite{ParticleDataGroup:2022pth,Colangelo:2022awx}. The LO relations among form factors obtained in \cite{Colangelo:2022awx} are extrapolated to the full kinematic range. }\label{Bc-fig1}
\end{center}
\end{figure}

The semileptonic $b \to c$ exclusive decays of $B_c$ to negative- and positive-parity charmonia are of particular interest, since the  matrix elements of the hadron currents $\bar c \Gamma b$ can be expressed near the zero-recoil point invoking the heavy quark spin symmetry \cite{Jenkins:1992nb,Colangelo:1999zn}. The consequence is that different processes can be related. The symmetry can be used to reconstruct the form factors of new physics operators starting from those computed, e.g., by lattice QCD \cite{Colangelo:2022lpy}. Moreover, $B_c$ decay processes provide us with  methods to investigate the nature of debated charmonia such as $X(3872)$ \cite{Ferretti:2013faa}, using observables in semileptonic  \cite{Colangelo:2022awx} and  nonleptonic channels \cite{Losacco:2023uvp}.  Due to the heavy quark spin symmetry, the four
$P-$wave charmonium states $\chi_{ci}$ (i=1,2,3) and $h_c$ belong to a spin 4-plet, and this holds also for the first radial excitations. If $X(3872)$ can be identified with $\chi_{c1}(2P)$, its  production  in semileptonic and nonleptonic $B_c$ modes would be precisely  correlated to the production of the other members of the charmonium multiplet in the same process;
such correlations  can be experimentally tested. An example is  in Fig.~\ref{Bc-fig1}, in which ratios of semileptonic decay distributions are plotted versus the variable $w=v \cdot v^\prime$, with $v$ and $v^\prime$ the four-velocities of $B_c$ and of the  produced charmonium state.  

\section{Example of a  BSM construction: the 331 model. Correlations among FCNC processes in down- and up-quark sectors}
Among the models constructed  enlarging the SM gauge group,  the  {\it 331  models} are based on  $SU(3)_c \times SU(3)_L \times U(1)_X$,   
  spontaneous broken first  to 
  $SU(3)_c \times SU(2)_L \times U(1)_Y$, then to $SU(3)_c \times U(1)_Q$  \cite{Singer:1980sw,Pisano:1991ee,Frampton:1992wt}.   Five additional gauge bosons and new fermions are in the spectrum:   the  left-handed SM fermions belong to triplets or antitriplets,   with the third component  usually a new heavy fermion.
In 331 models  the  anomaly cancelation and the  asymptotic freedom of QCD  constrain  the number of generations to be equal to the number of colours,  a rationale for such theories.
Moreover, the quark generations transform differently under  $SU(3)_L$, two generations  as triplets, one (usually the third generation)  as an antitriplet, a
difference that  can be invoked as  the origin of the large top quark  mass. 
 
 The relation  between the electric charge $Q$,  the $SU(3)$ generators  $T_3$ and $T_8$,  and the generator $X$ of $U(1)_X$:
 $  Q=T_3+\beta T_8+X $
introduces a parameter  $\beta$  defining the  variant of the model. For  $\beta$  multiple of $\displaystyle{\frac{1}{\sqrt{3}}} $ and of $\sqrt{3}$ the new gauge bosons $Y^{Q_Y^\pm}$ and $V^{Q_V^\pm}$  have integer  charges.
A  neutral gauge boson $Z^\prime$  mediates tree-level FCNC  in the quark sector, while the couplings to leptons are  diagonal and universal.  The extended Higgs sector  comprises  three $SU(3)_L$ triplets and one sextet.

\begin{figure}[t]
\begin{center}
\includegraphics[width = 0.9\textwidth]{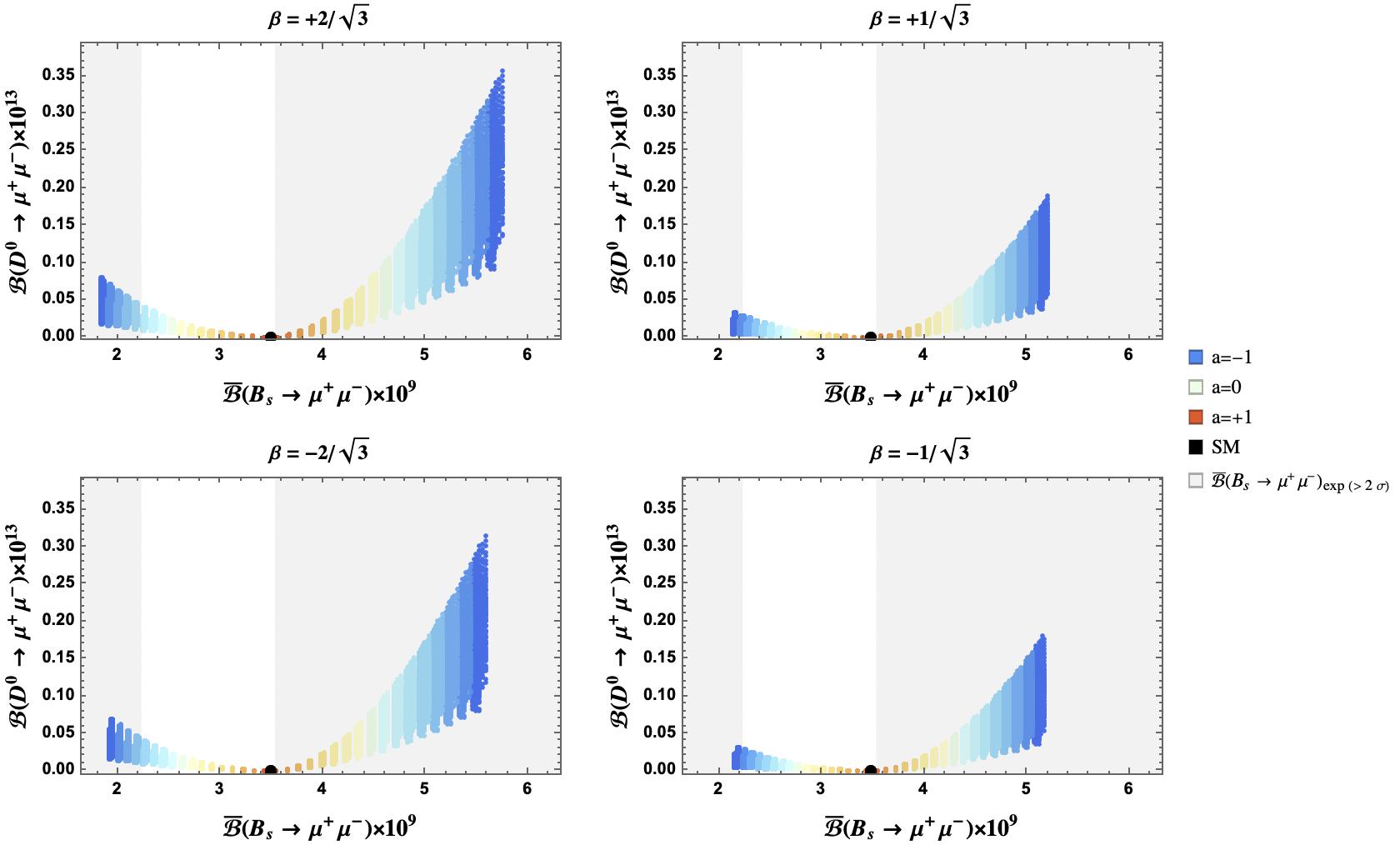}
    \caption{\baselineskip 12pt  \small Correlation between ${\cal B}(D^0 \to \mu^+ \mu^-)$ and $\overline {\cal B}(B_s \to \mu^+ \mu^-)$  in 331 models for a set of model parameters,  namely the parameter  $a$     entering in the  $Z-Z^\prime$ mixing \cite{Buras:2014yna}.
     The black dots are the SM result. The gray areas are to regions excluded by the  ${\cal {\overline B}}(B_s \to \mu^+ \mu^-)$ measurements  within 2$\sigma$  \cite{Buras:2021rdg}.}\label{corrD0Bs}
\end{center}
\end{figure}

In analogy with SM, the quark mass eigenstates are obtained  rotating the flavour eigenstates by  two unitary matrices, $U_L$ for up-type and $V_L$ for down-type quarks, with $V_{CKM}=U_L^\dagger V_L$. However, while in  SM $V_{CKM}$ only appears  in charged current interactions  and $U_L$ and $V_L$ never appear individually, in  331 models one can get rid  of only one matrix, either $U_L$ or $V_L$,  expressed in terms of $V_{CKM}$ and of the other matrix. The remaining rotation matrix enters in $Z^\prime$ couplings to quarks  \cite{Buras:2012dp}.  
Considering the  $Z^\prime$ interaction  with ordinary fermions,
correlations between observables in $B_{d,s}$ sectors and in the kaon sector can be established \cite{Buras:2012dp,Buras:2013dea,Buras:2014yna,Buras:2015kwd,Buras:2016dxz,Buras:2023ldz}.
For $\beta=\pm\displaystyle\frac{2}{\sqrt{3}}$ and $\beta=\pm\displaystyle\frac{1}{\sqrt{3}}$ and the third generation quarks in an antitriplet,   phenomenological constraints are satisfied, namely from $\Delta F=2$ observables in the $B_d,\,B_s, \,K$ systems and the electroweak precision observables, for  $Z^\prime$ in the  TeV range \cite{Buras:2013dea}. For  $\beta=\displaystyle\frac{2}{\sqrt{3}}$    relevant contributions to 
$\varepsilon^\prime/\varepsilon$ are predicted \cite{Buras:2015kwd}.  
The relation 
$ U_L=V_L \cdot V_{CKM}^\dagger $
induces  correlations between  FCNC transitions in the up- and down-type quarks,  a striking feature of the models. For example,  $c \to u \nu \bar \nu$ induced processes, e.g. $B_c \to B_u^{(*)} \nu {\bar \nu}$, can be related to $b \to s \nu {\bar \nu}$ and $s \to d \nu {\bar \nu}$  modes \cite{Colangelo:2021myn}.  
Fig.~\ref{corrD0Bs}  shows  the correlations  between  $D^0 \to \mu^+ \mu^-$  and $B_s \to \mu^+ \mu^-$  \cite{Buras:2021rdg}.

\section{Conclusions}
At present no undoubtable evidence is found of experimental deviations from SM: the above listed anomalies need to  be confirmed  with higher precision, with reduced hadronic uncertainty. The results of the new measurements and the arguments about the limitations of the theory will drive the analysis of the structure of fundamental interactions. 

\bigskip
\noindent{\bf Acknowledgments.}  We thank the organizers of HADRON 2023 for the invitation, and A.J. Buras, F. Giannuzzi, S. Nicotri and M. Novoa-Brunet for discussions.
The work has been carried out within the INFN projects (Iniziative Specifiche) QFT-HEP and SPIF.

\bibliographystyle{JHEP}
\bibliography{refhadron23}
\end{document}